\documentclass[useAMS,usenatbib]{mn2e}
\pdfoutput=1
\usepackage{lmodern}
\usepackage[utf8]{inputenc}
\usepackage[english]{babel}
\usepackage[T1]{fontenc}

\usepackage{savesym}
\savesymbol{arcmin}
\savesymbol{fs}
\savesymbol{arcsec}

\usepackage{rhmath}
\usepackage[nofloatbarriers]{rhgraphics}

\restoresymbol{}{arcmin}
\restoresymbol{}{fs}
\restoresymbol{}{arcsec}

\definecolor{darkblue}{HTML}{3333CC}
\definecolor{darkgreen}{HTML}{33CC33}

\DeclareMathOperator{\normcdf}{\Phi}
\DeclareMathOperator{\movingmedian}{\mathtt{movingmedian}}

\usepackage{tikz}
\usetikzlibrary{positioning,shapes,shadows,arrows}
\tikzstyle{decision} = [diamond, draw, fill=green!20, text width=4.5em, text badly centered, node distance=2.5cm, inner sep=0pt]
\tikzstyle{flownode} = [rectangle, draw, fill=blue!20, text width=5em, text centered, rounded corners, minimum height=2.5em]
\tikzstyle{line} = [draw, very thick, color=black!70, -latex']
\tikzstyle{cloud} = [draw, ellipse,fill=red!30, minimum height=2em]
\tikzstyle{block} = [text centered, text width=8em, minimum height=2em]

\usepackage[colorlinks=true,citecolor=blue,urlcolor=darkgreen]{hyperref}

\newcommand{\s}{\sin(2\pi\nu t_i)}
\renewcommand{\c}{\cos(2\pi\nu t_i)}
\renewcommand{\ss}{\sin^2(2\pi\nu t_i)}
\newcommand{\cc}{\cos^2(2\pi\nu t_i)}

\renewcommand{\uHz}{\ensuremath{\umu\mathrm{Hz}}}

\renewcommand{\eqref}[1]{Eq.~\ref{#1}}
\newcommand{\fref}[1]{Figure~\ref{#1}}
\newcommand{\code}[1]{\texttt{#1}}

\title[KASOC correction filter]{Automated preparation of \Kepler\ time series of planet hosts for asteroseismic analysis}
\author[R.~Handberg \& M.~N.~Lund]{R.~Handberg$^{1,2}$\thanks{E-mail: rasmush@bison.ph.bham.ac.uk} \& M.~N.~Lund$^{2,3,1}$ \\
$^1$ School of Physics and Astronomy, University of Birmingham, Edgbaston, Birmingham B15 2TT, United Kingdom\\
$^2$ Stellar Astrophysics Centre (SAC), Department of Physics and Astronomy, Aarhus University, DK-8000 Aarhus C, Denmark\\
$^3$ Sydney Institute for Astronomy (SIfA), School of Physics, University of Sydney, NSW 2006, Australia}
\date{Received 2014 September 1 / Accepted 2014 September 3}

\begin{document}
\maketitle

\begin{abstract}
One of the tasks of the Kepler Asteroseismic Science Operations Center (KASOC) is to provide asteroseismic analyses on Kepler Objects of Interest (KOIs).
However, asteroseismic analysis of planetary host stars presents some unique complications with respect to data preprocessing, compared to pure asteroseismic targets. If not accounted for, the presence of planetary transits in the photometric time series often greatly complicates or even hinders these asteroseismic analyses.
This drives the need for specialised methods of preprocessing data to make them suitable for asteroseismic analysis.
In this paper we present the KASOC Filter, which is used to automatically prepare data from the {\Kepler}/K2 mission for asteroseismic analyses of solar-like planet host stars.
The methods are very effective at removing unwanted signals of both instrumental and planetary origins and produce significantly cleaner photometric time series than the original data. The methods are automated and can therefore easily be applied to a large number of stars. The application of the filter is not restricted to planetary hosts, but can be applied to any solar-like or red giant stars observed by {\Kepler}/K2.
\end{abstract}

\begin{keywords}
methods: data analysis -- stars: oscillations (planet host -- transits)
\end{keywords}


\section{Introduction}\label{sec:intro}
The NASA \Kepler mission is dedicated to studying the distributions of extrasolar planets (exoplanets) and in particular Earth-like exoplanets.
Of particular interest is the characterisation of these exoplanets in terms of their habitability \citep[][]{2010Sci...327..977B}. 
The detection of exoplanets is achieved by virtue of the minute reduction of light as a planet moves in front of its host star, \ie, the transit method.
The transit signals themselves reveal only the relative size between the host star and the planet, so in order to infer the radius of the planet, one needs an independent measure of the host star radius. Furthermore, it is naturally of great importance to know the attributes of the host star (age, temperature, activity \etc) when studying the planetary system in general, and specifically for asserting the habitability.
One way to obtain this valuable information comes with an asteroseismic analysis of the signal imparted in the light curve from the pulsations of the host star.

Since the very start of the \Kepler mission it has been the task of the \emph{Kepler Asteroseismic Science Operations Center} \citep[KASOC;][]{2010AN....331..966K} to provide the asteroseismic community, in large represented by the \emph{Kepler Asteroseismic Science Consortium} (KASC), with \Kepler data, through the KASOC database\footnote{\url{http://kasoc.phys.au.dk/}}. The second main responsibility of KASOC is to provide asteroseismic analyses for stars suspected to host planets in order to determine stellar properties needed in the characterisation of the planetary systems. These targets are known as \emph{Kepler Objects of Interest} (KOIs) and the list of these is continuously updated as more and more data are analysed \citep[][]{2013ApJS..204...24B}.

\Kepler targets can be observed in two cadences, viz. \emph{Long Cadence} (LC; $\delta t\!\sim\!\SI{29.4}{min}$) and \emph{Short Cadence} (SC; $\delta t\!\sim\!\SI{58.89}{s}$), where $\delta t$ denotes the time between measurements. In the analysis of KOIs we are primarily interested in SC data as the maximum observable frequency, the \emph{Nyquist} frequency ($\nyquist\sim\SI{8500}{\uHz}$), is high enough for the detection of solar-like oscillations in main-sequence stars. But LC data are also useful for the studies of evolved Red Giant stars which are fully capable of hosting planetary systems \citep[see, \eg,][]{2013Sci...342..331H}.

Even though the photometric precision of \Kepler is very high \citep[][]{2010ApJ...713L..79K} and ideally suited for planetary studies as well as asteroseismology \citep[][]{2010PASP..122..131G}, the \Kepler data present (as do most space based instruments) a myriad of unique peculiarities and unwanted effects, primarily of an instrumental nature. These can complicate any analysis, be it of a planetary or asteroseismic signal. Some of these effects come in the form of \emph{outliers} in the time series, originating for instance from cosmic events, so-called \emph{Argabrightening} effects \citep[see, \eg,][]{VCleve},
 and momentum dumps; the variation in the heating of the space craft from the orbit around the Sun introduces \emph{drifts} from the induced focus change. Drifts can also come from changes in the pointing of the space craft, with the correction given by a so-called \emph{attitude tweak}. Temperature drifts are also often observed after times where the space craft has entered a \emph{safe mode}. \emph{Jumps} are seen in the time series between quarters due to the roll of the space craft, which shifts the position of the individual stars on the CCD focal plane, with each CCD having its own sensitivity response function. Furthermore, jumps are introduced after attitude tweaks or pixel sensitivity drops. Many of the effects perturbing \Kepler data are well-known and can thus be accounted for. One such effect is the signal from the LC sampling which produces peaks in the power spectrum of SC data at harmonics of $1/\mathrm{LC}$ \citep{2010ApJ...713L.160G}.
We refer the reader to \citet[][]{2010ApJ...713L..87J,Garcia2011,KeplerDataHandbook,VCleve} for more on the various instrumental effects found in the \Kepler data.

The \Kepler Mission Science Operations Center \citep[][]{2010ApJ...713L..87J,2010ApJ...713L.115H} produces data products optimised for the detection of planetary signals, and much goes into removing as many known artefacts as possible. However, these data are not necessarily the best for asteroseismic analysis and some problems have been reported for the various types of \emph{Pre-search Data Conditioning} (PDC) products, such as injected noise and perturbation of low frequency signals \citep[see, \eg,][]{2012MNRAS.422..665M,2013arXiv1307.4163G}. 

Much effort has already gone into the preparation of \Kepler data for general asteroseismic analysis \citep[see \eg][]{Garcia2011}. With some corrections being applicable to large samples of stars, while other are more targeted towards specific types of stars \citep[see \eg][]{2012AN....333..983J,2013arXiv1304.6673D}.
However, asteroseismic analysis of planetary host stars presents some unique complications compared to pure asteroseismic targets. In addition to noise originating from the previously mentioned effects, one also has to deal with the noise induced by the planetary companions. This is indeed a good example of the phrase ``\textit{In science, one man's signal is another man's noise}''\footnote{Edward Ng, New York Times, (1990) - a 20th century variation of Lucretius.}.

In this paper we describe the \emph{KASOC filter} which is specifically designed for the removal of the ``planetary noise'' in KOIs and preparation of the data for asteroseismic analysis. Moreover, the filter is able to correct data for stars without transit signature and can therefore be used in a more general sense for the data correction of solar-like and red giant stars. Eclipsing binary stars are subject to the same kind of noise as the planet hosts, with the distinction that the transit signatures here originate from a stellar companion. Due to this similarity the filter is also well suited to correct the data of eclipsing binaries.

The \Kepler space craft has now unfortunately lost two of its four reaction wheels, and consequently its ability to continue the hitherto ultra-fine pointing. The need for proper corrections of light curves for asteroseismic analysis will, however, not be any less as the majority of the KOIs still lack such an analysis, and as mentioned above the list of KOIs is ever increasing as more and more data are analysed for transit signatures by the \Kepler planet detection pipeline. Furthermore, with the planned continuation mission for \Kepler, dubbed ``K2'', data will keep coming and here likely new instrumental features will be introduced. K2 will not observe the \Kepler field towards the Cygnus-Lyra constellations, but instead monitor several fields along the ecliptic.

The structure of the paper is as follows: In \S\ref{sec:filter} we describe the components of the filter. The calculation of the power density spectrum is focus of \S\ref{sec:pspec}, while the performance of the filter is tested in \S\ref{sec:test} on a number of challenging cases. In \S\ref{sec:product} we describe the various data products that will be produced for the purpose of asteroseismic analysis. Finally, in \S\ref{sec:dis} we discuss our findings and look into potential future developments.


\section{The KASOC filter}\label{sec:filter}
The filter presented here is in fact an advancement of the initial KASOC filter, which has been in place at KASOC for several years \citep{oldkasocfilter}. While the methods described here are derived from ideas used in the original filter, the work presented here is a complete rewrite of the methods and the software\footnote{This work makes use of a number of open-source projects: Cython \citep{behnel2010cython}, matplotlib \citep{Hunter:2007}, \url{http://scipy.org}, \url{http://python.org}, \url{http://numpy.org}}.
Among the improvements compared to the previous methods, is the possibility to take known planetary periods into account, the details of which we will get back to later in this section, and the possibility of using it on LC data.

Having in mind that the filter is to be used on a large and diverse sample of stars, including planet hosts and eclipsing binaries, we set the following demands for our filter:
\begin{enumerate}[(i)]
\item It must be able to remove long-term trends and instrumental effects.
\item It must remove planetary signals -- both known and unknown.
\item It should not reduce the duty cycle for the observations or introduce a spectral window function.
\item It should work with both SC and LC data.
\item It must be simple.
\item It must be robust.
\item It must be fast.
\item It must be automated.
\end{enumerate}

The data type we use from \Kepler is the so-called \emph{raw target pixel data} \citep[][]{2010ApJ...713L..87J}.
These datasets consist of a series of digital images of the star in question, taken at consecutive timestamps, where no corrections have been made to the data, except simple corrections like cosmic-ray removal and background-subtraction. In the software developed to run the filter it is ensured that the time series is sorted in time, but this is only to speed up the following calculation.

In the following we will go through the working components of the filter one by one:

\paragraph*{Step 1.}
The very first step of the filter is the extraction of the light curve from the target pixel data.
The original \Kepler light curves are created from aperture photometry using pixel masks made to optimize signal-to-noise aimed for planet detection \citep[see, \eg,][]{2010ApJ...713L..97B,2011ApJS..197....6G}. However, this is not our aim, and from tests we find that modifying the original masks can yield significant improvements to the quality of the extracted light curves.
One such example is in relation to the thermal relaxations often seen after the \Kepler spacecraft has been turned towards Earth for data downloads or after safe-mode events.
As the temperature of the optics and camera settles the size and shape of the image of the stars change slightly. This causes structures in the extracted light curves resembling exponential decays.
In these cases, increasing the sizes of the masks can often remove this effect resulting in much cleaner time series. 

To redefine the pixel masks, we employ an iterative procedure developed by S.~Bloemen \etal\ (see Mathur \etal 2014, in prep.) which starts from the original \Kepler mask and iteratively adds or removes pixels to the edge of the aperture based on the amount of flux in each pixel. Pixels are added to the mask as long as they contain significant flux and as long as the flux is dropping when moving away from the center of the target. If on the other hand the flux starts rising, this is flagged as a contamination from a nearby bright star and the associated pixels are not used. The outcome of this is a new mask that most often is slightly bigger than the original mask. An obvious danger of this procedure is a slightly heightened risk of contamination from other faint nearby objects falling inside the aperture, but this risk is generally outweighed by the benefits. An exception could be dense regions in clusters.

Once the new mask is created, the light curve is extracted by simple aperture photometry, summing up all the light within the mask for each time step.
Our dataset from now on consists of a series of stellar brightness measurements, $x = \{x_1, x_2, \dots, x_N\}$, taken at incremental timestamps, $t = \{t_1, t_2, \dots, t_N\}$.

\paragraph*{Step 2.}
In the next step the extracted light curve is read-in and, as mentioned above, sorted in time. Timestamps with a flux value of -Inf is changed to NaN (Not-a-Number) -- this is purely for consistency with later steps in the filter. 
In addition we read in the values of the ``Quality'' flag. This entry contains bit values for each datum indicating if the specific datum has been subject to one or more known artefacts. See Table~\ref{tab:bits} in Appendix~\ref{sec:bit} for an overview of the different possible bit values \cite[see also][]{kepman} and our actions taken for each of these. For the ones with the ``Remove'' action, we set the flux value of the datum to NaN.

\paragraph*{Step 3.}
The filter now corrects the sudden jumps in measured flux caused by the roll of the spacecraft between observing quarters in addition to jumps at timestamps with a flagged discontinuity or attitude tweak in the original \Kepler data (\ie, having the action ``Correct'' in Table~\ref{tab:bits}). The end result of this is a relatively smooth concatenated time series where the individual observing quarters have been stitched together.

The changes between quarters can primarily be attributed to (multiplicative) sensitivity variations between the CCDs upon which a given star falls \citep[see, \eg,][]{2013ApJ...774L..19V,2013MNRAS.436.1576B}, see also \citet[][]{2012PASP..124..963K} and \citet[][]{Garcia2011}. Furthermore, additive changes between quarters from differences between pixel masks, and hence amount of captured flux, is kept at a minimum from the use of the generally larger custom masks, as described in Step~1 above. There might be a minor additive contribution from a change in the crowding of background stars.
Changes in flux arising from telescope attitude tweaks are additive as the cause here is a minor drift of the star on the CCD relative to the mask for the star. The attitude tweak repositions the star on its mask upon which a flux level close to the original is achieved.

\begin{figure}%
	\centering
	\includegraphics[width=\columnwidth]{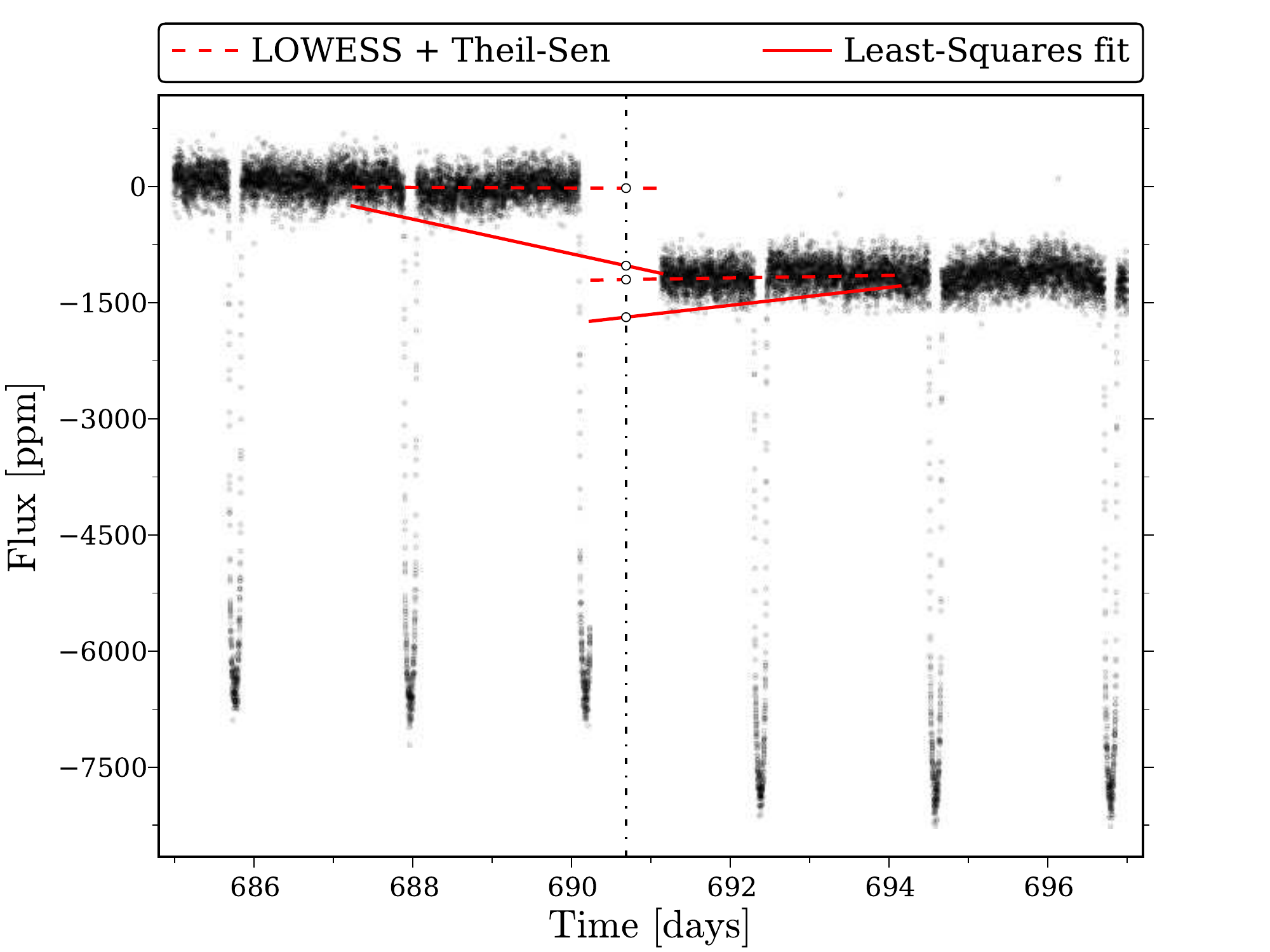}
	\caption{Illustration of the proposed method for obtaining the linear function for jump corrections (dashed red), namely by use of first a LOWESS smoothing followed by the Theil-Sen median slope estimation, here applied to a part of the HAT-P-7 time series. Also shown is the result of using a simple Least-Squares linear regression (solid red). The vertical dash-dotted line gives the time stamp at which the linear function are compared, with the applied correction given by the difference between the functions at this point.}%
	\label{fig:jump_corr}%
\end{figure}

We found that the following procedure for the stitching returned the best result with respect to the quality of the power spectrum, especially for planet hosts where transit features are removed (see \S\nobreak\ref{sec:hatp7} for more on the various procedures attempted):
Given a list of timestamps where jumps might be present, extracted from the flags or beginnings of quarters, constant offsets and linear trends are found in 3 day segments before and after each identified jump, and a correction is made to the part of the time series after the jump, either multiplicatively or additively, from the difference between the trends at the midpoint between the two sides of the jump.
The different models (no correction, constant offsets or linear trends) are then compared where the model with the lowest Bayesian Information Criterion (BIC) describing the 6 day segment around the jump is accepted and the correction is applied to the entire time series following the jump. Rather than making a least-squares (LS) fit of a linear function to the 3 day segments \citep[see, \eg,][]{2013MNRAS.436.1576B} we estimate the linear trend in a robust manner by first running a LOWESS\footnote{Using the Python module $\mathtt{statsmodels}$.} regression \citep[see][]{MR556476,lowess}, and on the smoothed signal find the trend as the Theil-Sen\footnote{Also known as the Kendall robust line-fit method.} median slope \citep[][]{MR0036489,MR0258201} see also \citet[][]{feigelson2012modern}. The added robustness over a LS fit is especially important for stars showing transits features, as the trend estimated for one side of a jump (\eg, between quarters) ending or starting in a transit will deviate greatly from the trend of the main out-of-transit component of the light curve (see \fref{fig:jump_corr}), and the correction will generally lead to a bad stitching.  

\paragraph*{Step 4.}
Next, gaps in the time series (\ie, missing measurements) are filled with NaNs, resulting in a time series on an (approximately) regular grid in time. This is done to ensure that in the following steps, a given number of data points will correspond to a constant span in time. It is important to emphasise that this does not create any new data points to be used, as the NaN values are ignored in all following steps, but it simply ensures that the filter behaves well at the edges of data gaps.

\paragraph*{Step 5.}
After these initial preparations of the time series, we have to take out the long-term trends arising both from instrumental drifts and stellar activity.
We do this by applying a moving median filter to the time series, essentially calculating the median of the measurements in a given window of time around each timestamp.
This generates a new low-passed version of the time series:
	\begin{equation}\label{eqn:xlong}
		x_\mathrm{long} = \movingmedian(x, \tau_\mathrm{long}) \, ,
	\end{equation}
where $\tau_\mathrm{long}$ is the time scale over which the median of data points are found. The default values for $\tau_\mathrm{long}$ is 3 days and 30 days for SC and LC respectively. Since the filter will remove any periodicities below $\tau_\mathrm{long}$, for any stars which have previously confirmed oscillations at low frequencies (Red Giants) $\tau_\mathrm{long}$ will automatically be scaled up to keep it significantly above the periods of the oscillations. This is also done for the KOIs where the known stellar parameters are used to estimate the expected frequencies of oscillations \citep[see, \eg,][]{Huber2013, Huber2014}, and these are then used to ensure that the filter does not perturb signals in that region.
The reason for choosing a moving median over, for example, the much faster moving average, is that the moving median is much more robust against outliers -- which is one of the criteria for the filter. This also means that it is much less influenced by any planetary transits and will follow the overall trends of the time series much better. See Appendix~\ref{sec:median} for details on the chosen implementation of the median filter. 

\paragraph*{Step 6.}
This step involves removal of any planetary features that are known \emph{a priori} to be present in the data. KASOC keeps a local database of known KOI parameters, which is routinely updated with parameters released by the \Kepler exoplanet team\footnote{\url{http://archive.stsci.edu/kepler/koi/search.php}}.
In fact we only use a minimum of information from these databases, namely the planetary orbital period, manifesting itself as the time between consecutive planetary transits. This parameter is one of the most basic parameters of the planetary transit light curve and does not require very sophisticated modelling of the planetary transits.

\begin{figure}%
	\centering
	\includegraphics[width=\columnwidth]{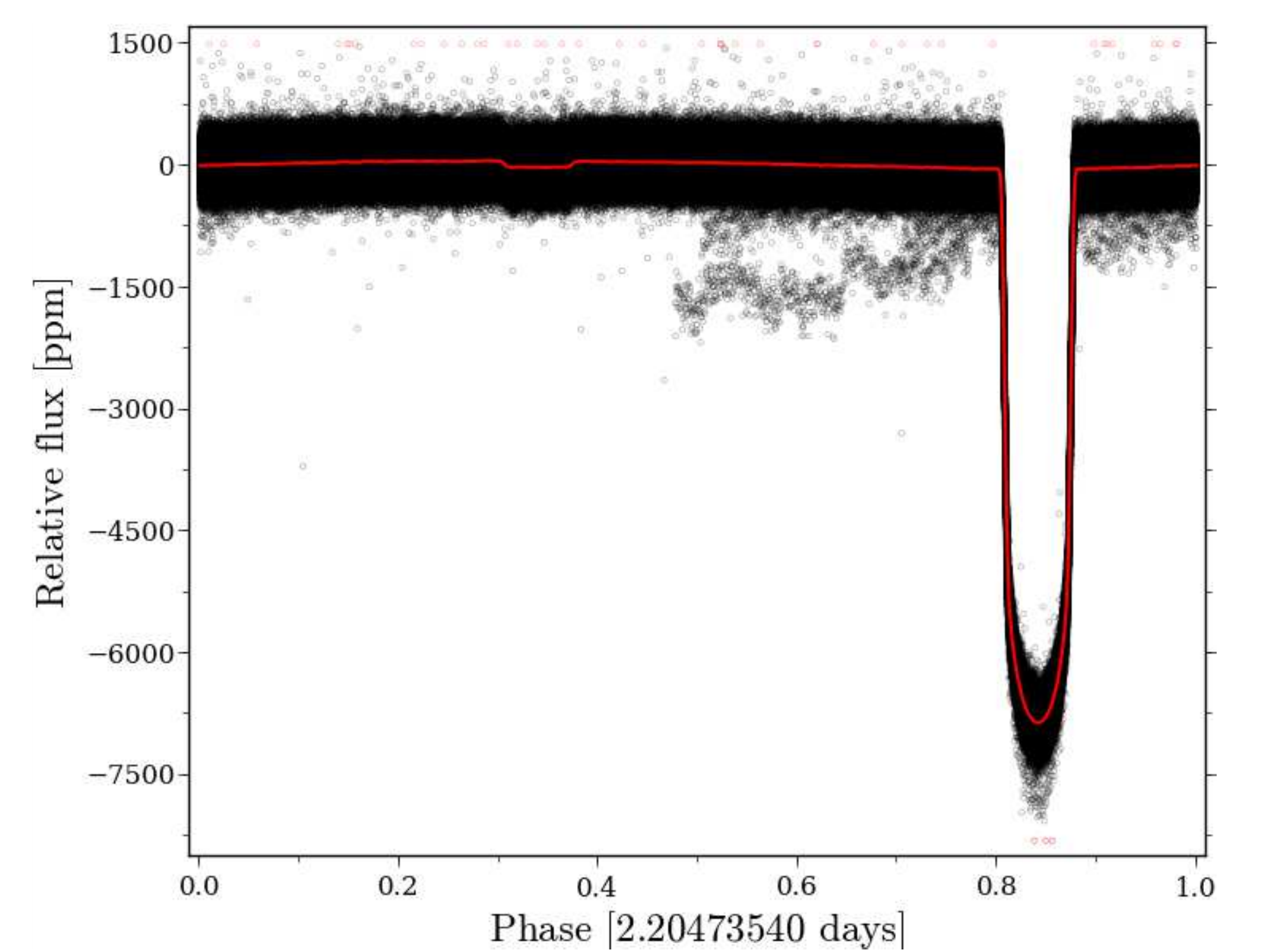}
	\caption{Phase curve of HAT-P-7 using the planetary period from \citet[][]{2013ApJ...774L..19V}. The phased data points are rendered as black points, while point with a relative flux value either above 1500 or below -8300 ppm have been truncated to these values and rendered as red points. The red curve gives the median filtered signal using a width of $P/1000$, which is used to correct the time series for transit features.}%
	\label{fig:phasecurve}%
\end{figure}

\begin{figure}
	\centering
	\begin{tikzpicture}[scale=2, node distance=1.5cm, auto]
		\node[cloud] (start) {Start};
		\node[flownode, below of=start] (start_n) {$n\!=\!1$\\$x_\mathrm{transit}\!=\!0$};
		\node[flownode, text width=15em, below of=start_n] (spectrum) {$\text{pc}(n)\!=\!\text{phasecurve}(x-x_\mathrm{transit}, n)$\\$x_\mathrm{transit}\!=\!x_\mathrm{transit} + \text{pc}(n)$};

		\node[flownode, below of=spectrum] (start_i) {$i\!=\!1$};
		\node[flownode, right of=start_i, node distance=3cm] (addn) {$n\!=\!n+1$};
		
		\node[block, below of=start_i, node distance=0.75cm] (arrow_anchor) {};
		\node[decision, below of=arrow_anchor, node distance=1.25cm] (condition_i) {$n\!=\!i$?};
		
		\node[decision, right of=condition_i, node distance=3cm] (end_condition) {$n\!=\!N_p$?};
		\node[cloud, below of=end_condition] (slut) {Final $x_\mathrm{transit}$};
		
		\node[flownode, text width=12em, below of=condition_i, node distance=2.5cm] (addfreq) {$x_\mathrm{transit}\!=\!x_\mathrm{transit} - \text{pc}(i)$};
		\node[flownode, text width=15em, below of=addfreq] (spectrum_two) {$\text{pc}(i)\!=\!\text{phasecurve}(x-x_\mathrm{transit}, i)$};
		\node[flownode, text width=12em, below of=spectrum_two] (removefreq_two) {$x_\mathrm{transit} = x_\mathrm{transit} + \text{pc}(i)$};
		\node[flownode, left of=addfreq, node distance=3cm] (addi) {$i\!=\!i+1$};
		\path[line] (start) -- (start_n);
		\path[line] (start_n) -- (spectrum);
		\path[line] (spectrum) -- (start_i);
		\path[line] (end_condition) -- node [near start, color=black] {Yes} (slut);
		\path[line] (end_condition) -- node [near start, color=black] {No} (addn);
		\path[line] (start_i) -- (condition_i);
		\path[line] (start_i) -- (condition_i);
		\path[line] (condition_i) -- node [near start, color=black] {Yes} (end_condition);
		\path[line] (addn) |- (spectrum);
		\path[line] (condition_i) --  node [near start, color=black] {No} (addfreq);
		\path[line] (addfreq) -- (spectrum_two);
		\path[line] (spectrum_two) -- (removefreq_two);
		\path[line] (removefreq_two) -| (addi);
		\path[line] (addi) |- (arrow_anchor.center);
	\end{tikzpicture}
	\caption{Flowchart showing the adopted strategy for the simultaneous ``fitting'' of phase curves for $N_p$ periods. The algorithm begins at ``Start'' and iteratively builds up $x_\mathrm{transit}$ by co-adding the phase curves for each period (upper loop), while each time taking previously calculated phase curves into account (lower loop). The final $x_\mathrm{transit}$ light curve is reached when all periods have been taken into account.}
	\label{fig:flowchart}
\end{figure}
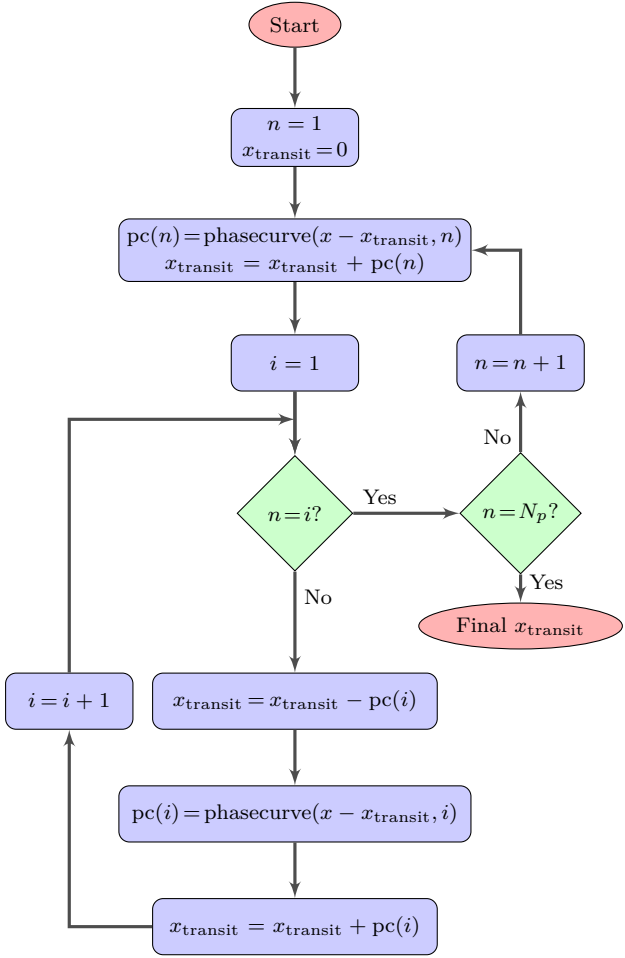

For each known planetary period, $P$, of a given KOI, we calculate the phase curve by folding the time series subtracted by the long term trend from \eqref{eqn:xlong}, \ie $x-x_\mathrm{long}$, with the period. The phase curve is then smoothed using a moving median filter with a width of $P/1000$ and subsequently a moving average of the same width, resulting in a robust triangular filter. Care is also taken that the smoothing of the phase curves are done in a cyclic way. This ensures that the resulting smoothed phase curve at phase 0 matches the value at phase 1. Once the smoothed phase curve has been obtained, it is unfolded back into the time domain to yield the transit light curve, $x_\mathrm{transit}$. An example of the phase curve of HAT-P-7b is shown in \fref{fig:phasecurve}. Note that the smoothed phase curve accurately follows both the primary transit (at phase $\sim\!0.85$), the secondary transit (at phase $\sim\!0.35$), and the modulation between the two caused by the changing phases of the planet \citep[see \eg][]{HATP7Science,2010ApJ...713L.145W}.
By working in phase space, the filter is only correlating data points that are well separated in time and we are therefore able to perform the smoothing in a quite narrow window, but at the same time minimising the risk of introducing unwanted signals into the time series.
In cases of multiple planet systems \citep[see \eg][]{Kepler36,Kepler37,Kepler5065,Kepler56}, phase curves are calculated iteratively for each known planetary period in a way that ensures that the different planets have minimal influence on each others phase curves. The exact procedure is shown in \fref{fig:flowchart} and, in short, consists of redoing all previously derived phase curves whenever a phase curve has been calculated.

\paragraph*{Step 7.}
The filter now goes on to removing any sharp features still left in the time series that are not accounted for by $x_\mathrm{long}$ and $x_\mathrm{transit}$. This should remove both instrumental effects like sudden jumps in flux level from pixel sensitivity drop outs, cooling after safe-modes \citep[see][]{KeplerDataHandbook} or other sharp features. In particular it should also remove transit features that are not yet recorded.
This is accomplished by running another moving median filter with a short time-scale through the time series where we have first subtracted $x_\mathrm{long}$ and $x_\mathrm{transit}$. The new time series is constructed as follows:
	\begin{align}\label{eqn:xshort}
		x_\mathrm{short} =& \movingmedian(x-x_\mathrm{long}-x_\mathrm{transit}, \tau_\mathrm{short}) \nonumber \\
		&+ x_\mathrm{long} + x_\mathrm{transit} \, ,
	\end{align}
where $\tau_\mathrm{short}$ is the chosen time scale of the short filter. The default values of $\tau_\mathrm{short}$ is 1 hour and 0.5 days for SC and LC respectively. To detect the presence of sharp features we construct the following diagnostic signal:
\begin{equation}\label{eq:w}
		w = \frac{x_\mathrm{long} + x_\mathrm{transit}}{x_\mathrm{short}} - 1 \, .
	\end{equation}
When the different filters follow each other closely the diagnostic signal will be close to zero, but if a sharp feature is present the short filter will deviate from the long and the diagnostic signal will in turn also deviate from zero.
	
\paragraph*{Step 8.}
The relative standard deviation of this signal, $\sigma_w/\mean{\sigma_w}$, is then calculated using the moving \emph{median absolute deviation} (MAD).
The rationale is now that when this quantity is large, meaning that the short filter deviates a lot from the long filter, we should give higher weight to the short filter. In other words, we create the final filter as a weighted mean between the short and long filters:
	\begin{equation}\label{eq:filter}
		\mathrm{filter} = c \cdot x_\mathrm{short} + (1-c) \cdot (x_\mathrm{long} + x_\mathrm{transit}) \, ,
	\end{equation}
where $c$ is the \emph{turnover function} between the two filters returning values between 0 and 1, given $\sigma_w/\langle\sigma_w \rangle$. In practice any function with these characteristics can be used, but we have opted for using the cumulative normal distribution function, $\normcdf(x;\mu,\sigma)$:
	\begin{equation}\label{eq:turnover}
		c = \normcdf(\sigma_w/\mean{\sigma_w}; \mu_\mathrm{to}, \sigma_\mathrm{to}) \, ,
	\end{equation}
where $\mu_\mathrm{to}\!=\!5$ and $\sigma_\mathrm{to}\!=\!1$ is used as the default values. The value of $\mu_\mathrm{to}$ is the value at which the two filters are equally weighted ($c\!=\!0.5$) and $\sigma_\mathrm{to}$ indicate the smoothness of the transition between the two. In the special case of $\sigma_\mathrm{to}\!=\!0$ the Heaviside step function is used instead.

\paragraph*{Step 9.}
Once the filter function of \eqref{eq:filter} is created the original time series, $x$, is corrected by dividing with the filter function, yielding a new filtered time series, $x_\mathrm{filt}$, which contains relative flux measurements in parts-per-million (ppm): 
	\begin{equation}\label{eq:filtering}
		x_\mathrm{filt} = 10^6 \left( { \frac{x}{\mathrm{filter}} - 1 } \right) \, .
	\end{equation}

\paragraph*{Step 10.}
The last step of the filter is to estimate the error on the measurements as the internal scatter in the time series. Since we have already removed the trends, the standard deviation can be robustly estimated by calculating the MAD (which now can be found simply as the moving median of the absolute value of the filtered signal) and scaling it to yield the standard deviation on the $\tau_\mathrm{long}$ time scale:
	\begin{equation}\label{eq:sigma}
		\sigma = k \cdot \movingmedian(|x_\mathrm{filt}|, \tau_\mathrm{long}) \, .
	\end{equation}
The conversion factor from MAD to standard deviation is $k \equiv 1/\p*{ \normcdf^{-1}(\tfrac{3}{4})}\approx1.4826$. Sigma-clipping is performed in which any points where $|x_{\mathrm{filt},i}| > 4.5\sigma_i$ are removed from the time series (set to NaN). To avoid any contamination from bad data points, the last step should be repeated iteratively until no more points are removed. However, in practice this has little implication and is omitted in order to speed up the calculation.

The end products of the filter are therefore a new set of filtered data points $\{x_{\mathrm{ filt},i}\}$ and estimated errors, $\{\sigma_i\}$, associated with each datum.

The spectral response of the final filter is near impossible to quantify in detail, since it not only depends on the specific settings that the filter is run with, but also on the quality of the actual data on which it acts. However, we do provide all information required in order to investigate this for each individual target in the provided data products (see~\S\ref{sec:product}).
In general terms, the spectral response will be dominated by a $\sinc^2(\nu)$-like behaviour, arising from $x_\mathrm{long}$, which will remove power at low frequencies. Frequencies below $1/\tau_\mathrm{long}$ should therefore be used with care.
Depending on the presence of planet transits or sudden jumps accounted for by $x_\mathrm{transit}$ and $x_\mathrm{short}$ respectively, the filter will also influence at higher frequencies, \eg removing harmonics of the planetary period.


\section{Power density spectra}\label{sec:pspec}
In addition to the corrected time series, KASOC also produce two versions of power density spectra of the corrected time series.
These power spectra are the data products that in most cases will be used for detection and characterisation of the stellar oscillations \citep[see, \eg,][]{Huber,Hekker,Savita,2011arXiv1104.0631V,2011A&A...527A..56H,2012MNRAS.427.1784L}.

\paragraph*{Lomb-Scargle} The first version is a classical Lomb-Scargle periodiogram \citep{Lomb1976,Scargle1989} using the fast implementation of \citet{PressRybicki1989}. The spectrum is calculated at critical sampling and with frequencies from 0 to $\nyquist\approx1/(2\cdot\delta t)$, where $\delta t$ is the median time sampling, calculated as the median of the pairwise difference between the timestamps.
Normalisation is done such that the total power in the positive half of the power spectrum satisfies Parseval's theorem:
	\begin{equation}
		\sum_{j=1}^{N/2} P_j = \frac{1}{N} \sum_{i=1}^N (x_i - \mean{x})^2
	\end{equation}

\paragraph*{Weighted Least Squares spectrum} The second version of the power spectra are calculated based on weighted least squares sine wave fitting to the time series \citep[see \eg][]{KjeldsenThesis,Frandsen1995}. The residuals of the weighted least squares fit for a given frequency, $\nu$, can be written as:
	\begin{equation}
		R(\nu) = \sum_{i=1}^{N} w_i \tuborg*{ x_i - \kant*{\alpha\cdot\sin(2\pi\nu t_i) + \beta\cdot\cos(2\pi\nu t_i)} }^2 \, ,
	\end{equation}
where $w_i$ denotes the statistical weight of each point, usually chosen as $w_i=1/\sigma_i^2$.
Minimising the above expression with respect to $\alpha$ and $\beta$ gives the following solutions:
	\begin{align}
		\alpha(\nu) &= \frac{s \cdot cc - c \cdot sc}{ss \cdot cc - sc^2} \label{eqn:alpha}\, , \\
		\beta(\nu) &= \frac{c \cdot ss - s \cdot sc}{ss \cdot cc - sc^2} \label{eqn:beta}\, ,
	\end{align}
where we have introduced the following short-hands for the various sums:
	\begin{align}\begin{split}\label{eqn:summer}
		s &\equiv \sum\nolimits_i w_i \cdot x_i \cdot \s \\
		c &\equiv \sum\nolimits_i w_i \cdot x_i \cdot \c \\
		ss &\equiv \sum\nolimits_i w_i \cdot \ss \\
		cc &\equiv \sum\nolimits_i w_i \cdot \cc \\
		sc &\equiv \sum\nolimits_i w_i \cdot \s \cdot \c
	\end{split}\end{align}
The weighted power density spectrum can then be calculated as follows:
	\begin{equation}\label{eqn:PLS}
		P(\nu) = \frac{\Delta T}{2} \kant*{ \alpha(\nu)^2 + \beta(\nu)^2 } \, ,
	\end{equation}
where $\Delta T$ is the \emph{effective} observation length, taking gaps in the time series into account. Normalising with $\Delta T/2$ has the effect that a sine wave with an amplitude of $a$ will have a peak with an area of $\tfrac{a}{2}$ in the power spectrum, or what is also known as the rms-scaled power density. The value of $\Delta T$ can be found from the spectral window as given below in \S\nobreak\ref{sec:specwin}.

The final procedure that we use for generating the weighted power density spectra is as follows:

\begin{enumerate}[(i)]
	\item Estimate the critical sampling as $\delta\nu=1/(t_N-t_1)$ 
		  and the Nyquist frequency as $\nyquist=1/(2\cdot\delta t)$.
	\item Obtain $\Delta T$ (\eqref{eqn:DeltaT}) from the spectral window function.
	\item Calculate the power density spectrum (\eqref{eqn:PLS}) from $\nu=0$ to $\nyquist$ with a critical sampling of $\delta\nu=1/\Delta T$.
\end{enumerate}


\subsection{Spectral window}\label{sec:specwin}
The spectral window is needed in the conversion between a power spectrum in units of $\mathrm{ppm}^2$ to one in units of power density, \ie $\mathrm{ppm}^2/\mathrm{\mu Hz}$. Furthermore, the spectral window can be used in detailed \emph{peak-bagging} efforts, as the observed power spectrum will be given by the underlying stellar spectrum convolved by the spectral window function.
The spectral window function can be calculated by utilizing the weighted least squares spectrum (Eqs.~\ref{eqn:alpha} and \ref{eqn:beta}) from above:
	\begin{equation}\label{eqn:Pwin}\begin{split}
		P_\mathrm{win}(\nu; \nu_w) = \tfrac{1}{2} \big\{& \alpha_\mathrm{sin}(\nu_w+\nu)^2 + \beta_\mathrm{sin}(\nu_w+\nu)^2 \\ &+ \alpha_\mathrm{cos}(\nu_w+\nu)^2 + \beta_\mathrm{cos}(\nu_w+\nu)^2 \big\} \, ,
	\end{split}\end{equation}
where $x_i\!=\!\sin(2\pi\nu_w t_i)$ (for $\alpha_\mathrm{sin}$ and $\beta_\mathrm{sin}$) and $x_i\!=\!\cos(2\pi\nu_w t_i)$ (for $\alpha_\mathrm{cos}$ and $\beta_\mathrm{cos}$). Here $\nu_w$ is a dummy frequency which we set to $\nyquist/2$.

As mentioned above $\Delta T$ is the \emph{effective} observation length, taking gaps in the time series into account.
 The effective observation length can be calculated as the inverse of the area under the spectral window function:
	\begin{equation}\label{eqn:DeltaT}
		\Delta T = \kant*{ \int_{-\infty}^{\infty} P_\mathrm{win}(\nu^\prime; \nu_w) \, d\nu^\prime }^{-1} \, .
	\end{equation}
The following steps will be taken for the calculation of the spectral window:
\begin{enumerate}[(i)]
	\item Estimate the critical sampling as $\delta\nu=1/(t_N-t_1)$ 
		  and the Nyquist frequency as $\nyquist=1/(2\cdot\delta t)$.
	\item Calculate $P_\mathrm{win}(\nu; \nyquist/2)$ with a sampling of $\delta\nu/10$ from $\rm -300\, \mu Hz$ to $\rm 300\, \mu Hz$ (\eqref{eqn:Pwin}).
	\item Integrate the window function using trapezoidal integration to yield $\Delta T$ (\eqref{eqn:DeltaT}).
\end{enumerate}

\section{Tests}\label{sec:test}
In the following we test the performance of the filter on a number of challenging cases. The test cases comprise stars with both strong and weak planetary signal and artificial time series. All data was downloaded from the KASOC database.


\subsection{HAT-P-7 (Deep transits)}\label{sec:hatp7}
HAT-P-7 (KOI-2; KIC~10666592; Kepler-2) is one of the very well-studied planetary systems also observed with \Kepler \citep[see, \eg,][Lund \etal 2014 (submitted), Benomar \etal 2014 (submitted)]{2008ApJ...680.1450P,2009ApJ...703L..99W,2010ApJ...713L.164C}. We chose this star because it has very pronounced transit features with a dimming of about ${\sim}7000\,\rm ppm$ from its close-in hot-jupiter planet HAT-P-7b, and thus presents a prime case for the test of the removal of transit signal.

\begin{figure*}%
	\centering
	\includegraphics[width=2.3\columnwidth]{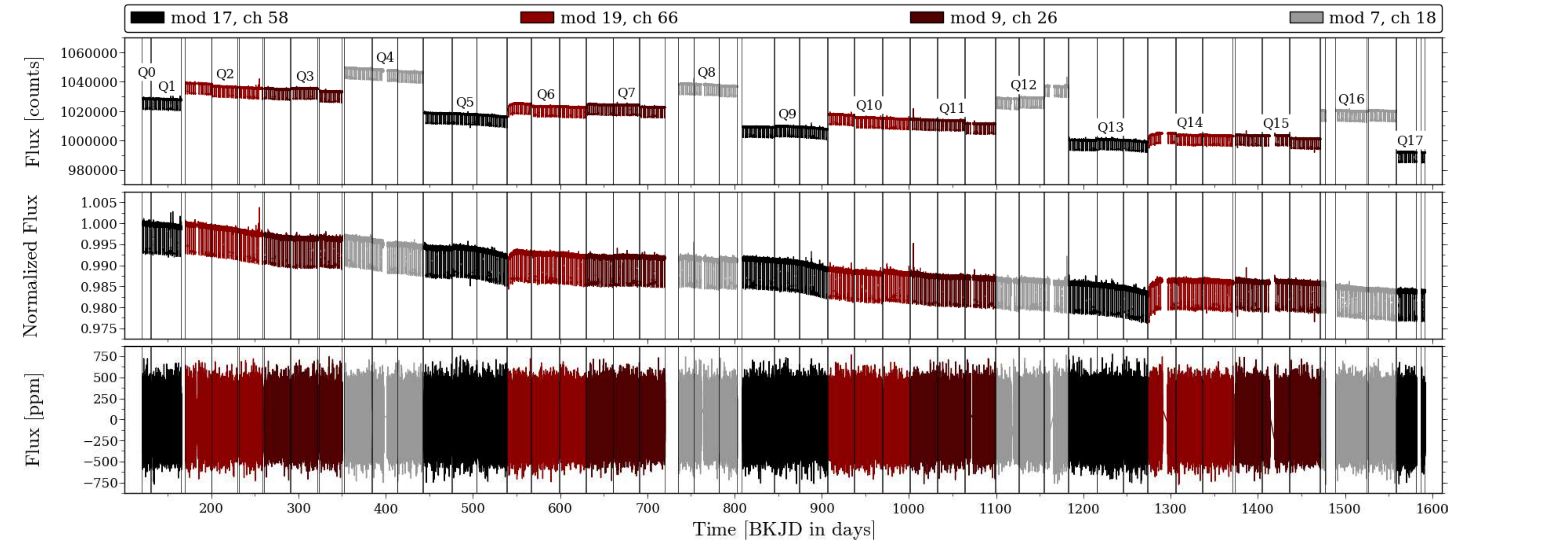}
	\caption{Kepler time series for HAT-P-7. \textbf{Top:} Initial time series after removal of bad data points. Indicated is the quarter designation with the vertical lines giving the start and end times for the sub-quarters. The different colouring gives the position of HAT-P-7 with respect to the \Kepler CCD module (mod) and channel (ch). \textbf{Middle:} Stitched light curve from following the procedure outlined in Step~3 of \S\nobreak\ref{sec:filter}. \textbf{Bottom:} Final corrected light curve ready of asteroseismic analysis after applying all steps of \S\nobreak\ref{sec:filter}.}%
	\label{fig:timeseries}%
\end{figure*}
  
For the phase-folding we adopted the period of ${P=2.20473540}$ days from \citet[][]{2013ApJ...774L..19V}. The resulting phase-folding light curve for HAT-P-7 has already been presented in \fref{fig:phasecurve}. The time series of HAT-P-7 extracted from pixel-data can be seen in the top panel of \fref{fig:timeseries} after Steps 1 and 2 from \S\nobreak\ref{sec:filter}. We have indicated the start and end times of each quarter (vertical lines) in addition to the specific modules (mod) and channels (ch) of the \Kepler CCD panel that HAT-P-7 falls on. From this our assumption of CCD sensitivity differences as the cause for the quarter-to-quarter jumps is substantiated as the same trend is seen within each \Kepler year, \eg does module 7 always have the highest flux level while module 17 has the lowest. The overall decrease in flux level during the approximately four \Kepler years of observation can possibly be attributed to the degradation of the instruments.  
The middle panel gives the stitched light curve from Step~3 in \S\nobreak\ref{sec:filter}. This light curve is very continuous and only some strong bends at the beginning of quarters 6 and 14 breaks the overall decrease of the light curve as the linear trend found here is instrumental - likely thermal trends. The bottom panel gives the final corrected light curve. In the power spectrum of this light curve (not shown) virtually no power is left from residuals of the planetary signal. These would have been seen at integer values of the frequency of the planetary period. We also tried a stitching where all jump corrections are made additively by comparing the median levels in 1 day segments before and after the jump and then correcting the suspected jump by adding the difference in median levels to the part of the time series after the jump. This is the same procedure as used in \citet[][]{Garcia2011}, with the difference that these authors instead use the mean levels in 1 day segments. In this purely additive approach clear residuals from the planet signal was present in the low-frequency part of the power spectrum. At higher frequencies only small differences were seen between the power spectra, but the stitching procedure used here generally gave lower power values which could affect mode amplitudes and widths estimated in an asteroseismic analysis. The reason for the planetary signature in the additive-only approach can likely be attributed to a larger scatter in the phase curve used for the correction of the transits because different parts of the time series has different scales and hence different transit depths.


\subsection{Kepler-22 (Weak transit signal)}
\begin{figure*}%
	\centering
	\includegraphics[width=2.2\columnwidth]{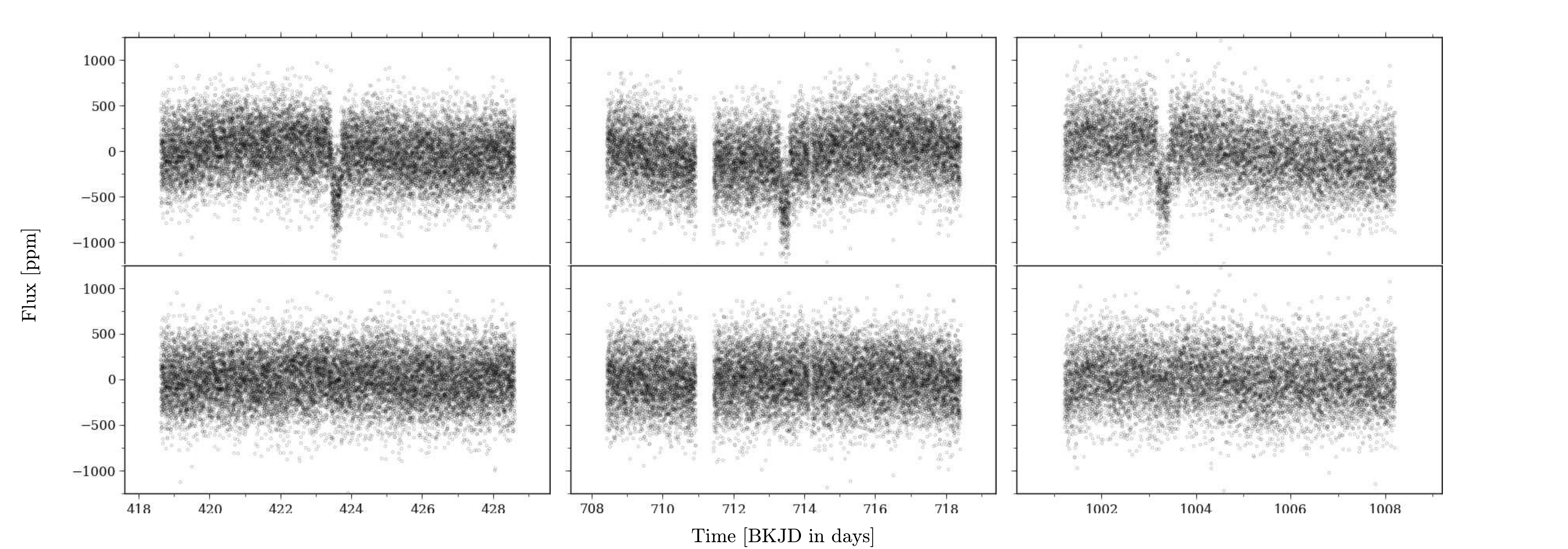}
	\caption{Planetary transit signal in SC data of Kepler-22 before (top panels) and after (bottom panels) the application of the filter, assuming no a priori knowledge of the period.}%
	\label{fig:kepler22}%
\end{figure*}

Kepler-22 (KIC~10593626; KOI-87) was reported by \citet{Kepler22b} as the first confirmed planet within the habitable zone of its parent star. This means that it could potentially harbour life on its (possibly rocky) surface. 
The planet has a period of ${\sim} 290$ days and the planet has a radius a couple of times that of Earth.
We have selected this target as a test case to demonstrate the ability of the filter to remove planetary transits even without any prior knowledge of the planetary period. We ran the raw data of Kepler-22 through our filter, without specifying any planetary period and thereby omitting Step~6 of \S\nobreak\ref{sec:filter}. In \fref{fig:kepler22} we show in the top panels segments of the time series around three of the four transits covered by the SC data, and this after running the filter up to and including Step~4. In LC data one additional transit can be found \citep[see][]{Kepler22b}. Unfortunately the transit at $\rm BKJD\sim1293$ days happens to fall directly within a gap in the data. The lower panels show the outcome of running the filter to end. The filter successfully removes the small planetary transits of only ${\sim}500\,\rm ppm$, which should be held against the general spread in the time series which is of similar size, leaving the time series clean and ready for asteroseismic analysis. Such an asteroseismic analysis was performed in \citet{Kepler22b}, albeit only including 19 (Q2.3-Q8) months of SC data and yielding only a marginal detection of the $p$-mode power excess. Currently, 42 months of SC data (Q2.3-Q9.3 + Q11.1-Q17.2) are available which together with a better correction of the time series should yield a more firm detection of $p$-modes and with it a better understanding of Kepler-22 and its planet -- such an updated analysis is currently under preparation (T. Arentoft, private communication, 2014).    


\subsection{Kepler-65 (multi-planet system)}
Kepler-65 (KOI-85; KIC~5866724) is a triple planet system observer by \Kepler in SC for 27 months (Q3 through Q11).
The three planets all have short periods and similar sizes (1.42, 2.58 and 1.52 $\mathrm{R}_\oplus$), making the system a good test case for our removal of multiple planets.

\begin{figure}%
	\centering
	\includegraphics[width=\columnwidth]{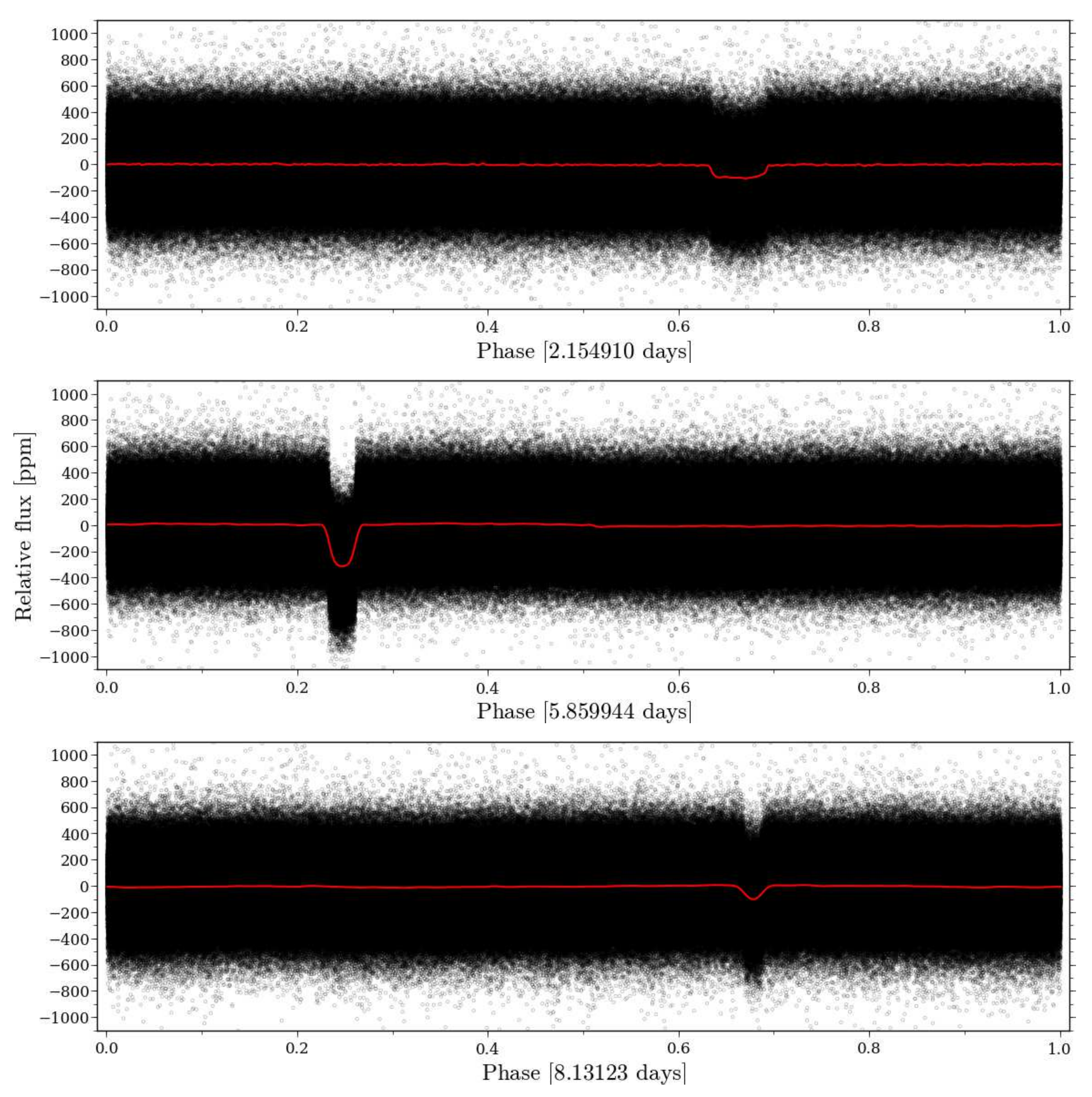}
	\caption{Kepler-65 phase curves. The panels show the phase curves from the iterative scheme used for multiple-planet systems, one for each of the three planets identified in the system with known periods from \citet[][]{Kepler5065}.}%
	\label{fig:kepler65}%
\end{figure}

The data was passed through the filter using the default settings and the known planetary periods from \citet[][]{Kepler5065}.
In \fref{fig:kepler65} the phase curves for the three planets as returned by the filter are shown.
Note that the three extracted smoothed phase curves are very clean and do not seem perturbed by the presence of the other planets as a consequence of our iterative smoothing routine.
 
\begin{figure}%
	\centering
	\includegraphics[width=\columnwidth]{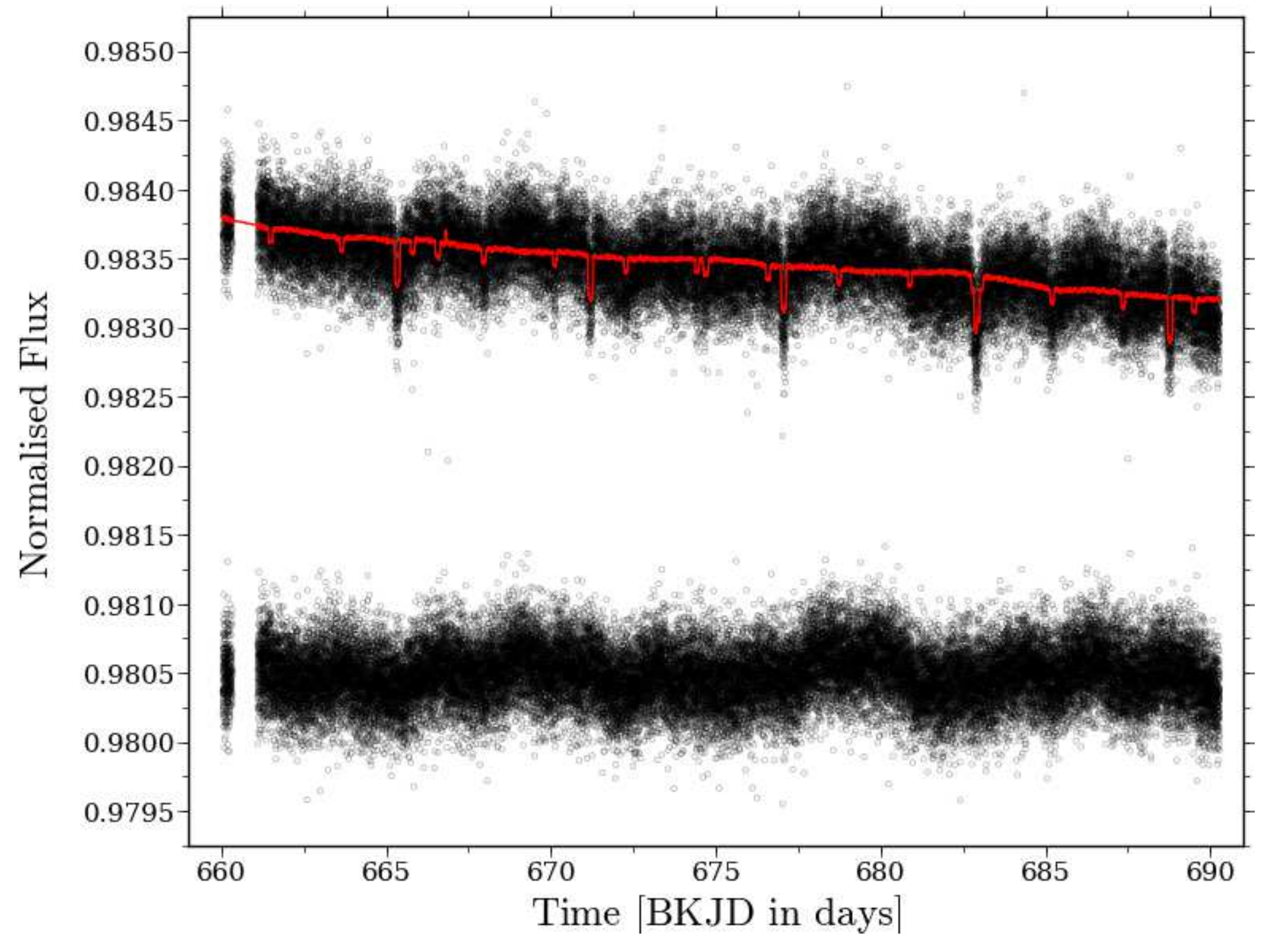}
	\caption{Filter applied to Kepler-65 SC data, here on the Q7.2 1-month segment. The top curve shows in black the time series after the stitching and over-plotted in red with the filter (Step~8, \eqref{eq:filter}) including the iterative inclusion of the three planetary transits (see Figures~\ref{fig:flowchart} and \ref{fig:kepler65}). The bottom curve shows the time series after the division of the filter and sigma-clipping. For clarity we have offset the corrected curve and omitted conversion to ppm.}%
	\label{fig:kepler65filt}%
\end{figure}

As previously described, the phase curves are then combined and ``unfolded'' into the time series to yield a full combined transit curve.
This is shown in \fref{fig:kepler65filt} together with the time series before and after the filter is applied.
As it can be seen, the complex transit signatures are very effectively removed, leaving a clean time series without traces of any of the planetary transits.
We note that the autocorrelation of this corrected time series shows a clear signal at around a period of ${\sim}8$ days, which is also the rotation period estimated in \citet[][]{Kepler5065}.


\subsection{KIC~5023732 (Red giant)}
KIC~5023732 is an evolved Red Giant star observed by \Kepler in LC throughout the mission (Q0--Q17).
It is a Red Giant branch star that is a member of the open cluster NGC~6819 \citep{Stello2011,Corsaro2012}. It exhibits solar-like oscillations around $\numax=27.1\pm0.2\,\rm \mu Hz$, and here serves the purposes, as a randomly chosen target, to demonstrate the KASOC filters performance on long cadence data of a star with relatively low oscillation frequencies.

\begin{figure}
	\centering
	\includegraphics[width=\columnwidth]{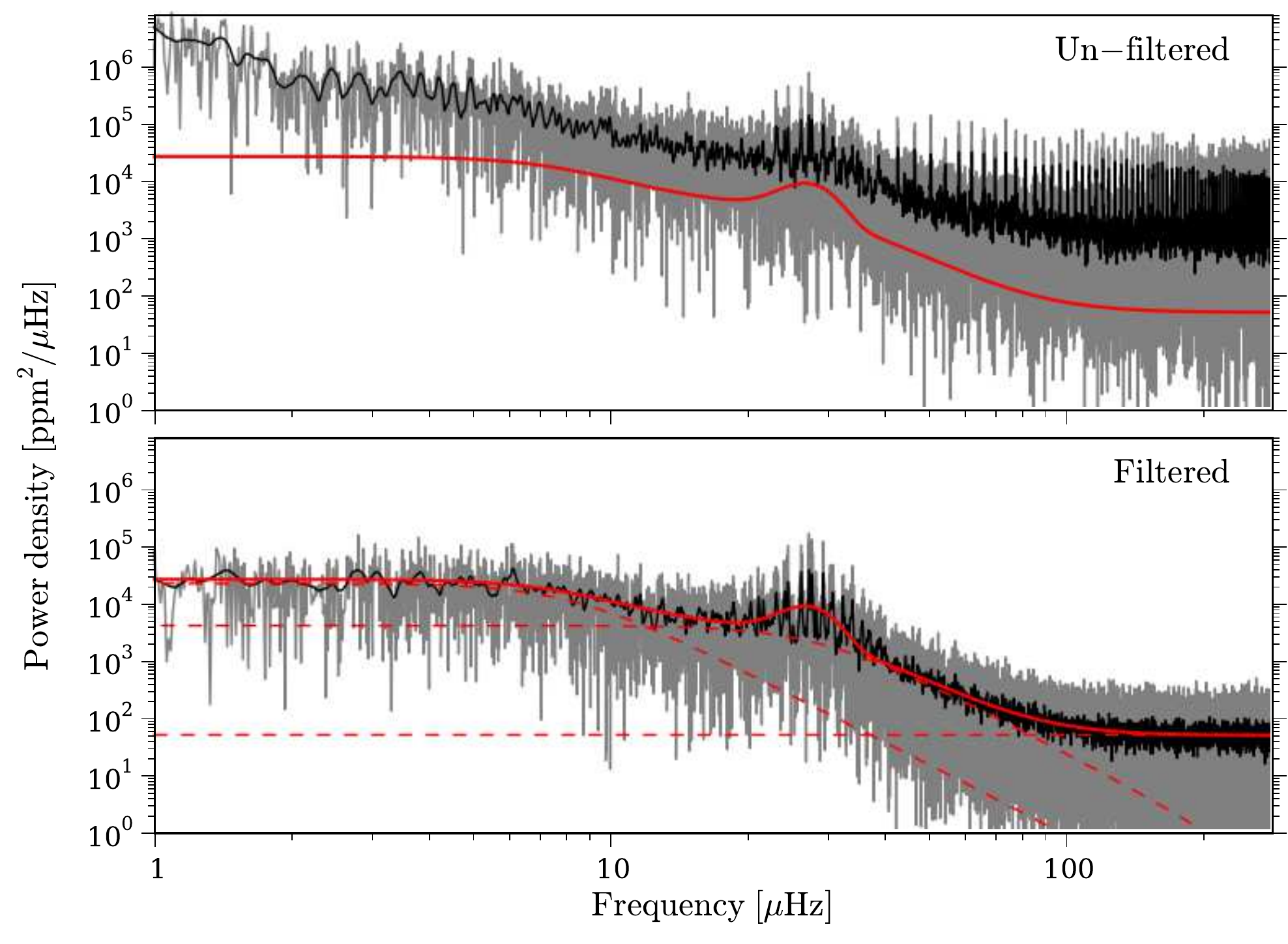}
	\caption{Power density spectrum of the red giant star KIC~5023732 (grey) before and after the filter has been applied. In black the spectrum is shown after a $\rm 0.25\ {\mu}Hz$ smoothing. The filtered spectrum is very clean and oscillation modes are clearly visible. In the bottom panel the solid red line indicates a fit to the expected shape of the power spectrum and the individual components are shown as dashed lines. The background is also shown in the top panel for comparison.}
	\label{fig:redgiant}
\end{figure}

The filter was run in the standard configuration, and the resulting power density spectrum of the corrected time series is shown in \fref{fig:redgiant}.
From the spectrum it is clearly seen that the star has solar-like oscillation modes sitting on top of a background that would be attributed to noise from the granulation in the outer layers of the star, which in turn is driving the oscillations. At high frequencies the spectrum becomes completely flat as it is here dominated by the photon shot noise.

In the lower panel of \fref{fig:redgiant}, the solid red line indicates a fit to a standard description of the stellar background signal composed by two Harvey-like profiles accounting for the granulation and faculae signals plus a white background for the photon shot noise \citep[see \eg][]{Harvey1985,Aigrain2004,Kallinger2014}. The oscillations are described as a Gaussian envelope. The full model of the power spectrum therefore becomes:
	\begin{align}\begin{split}\label{eqn:background}
		\mathscr{P}(\nu) = W &+ \eta(\nu) \sum_{i=1}^2 \frac{4\sigma_i^2\tau_i}{1 + (2\pi\nu\tau_i)^4} \\
		                     &+ \eta(\nu)\, a_\mathrm{env} \exp\kant*{\frac{-(\nu - \numax)^2}{2c_\mathrm{env}^2}} \, ,
	\end{split}\end{align}
where $\eta(\nu)\equiv\sinc^2(\nu\delta t)$ is the attenuation of signals arising from the finite exposure time, $W$ is the white noise level and $\sigma_i$ and $\tau_i$ indicates the amplitudes and timescales of the different components.
As it can be seen from this, the filter is able to handle long cadence red giant stars very well and produce very clean power spectra dominated by signals of stellar origins. This is sustained by the fact that the resulting filtered power spectrum very closely follows the shape we expect for a red giant stars (\eqref{eqn:background}) across the entire frequency range.

\subsection{16 Cyg A (Artificial transit light curve)}
\begin{figure}%
	\centering
	\includegraphics[width=\columnwidth]{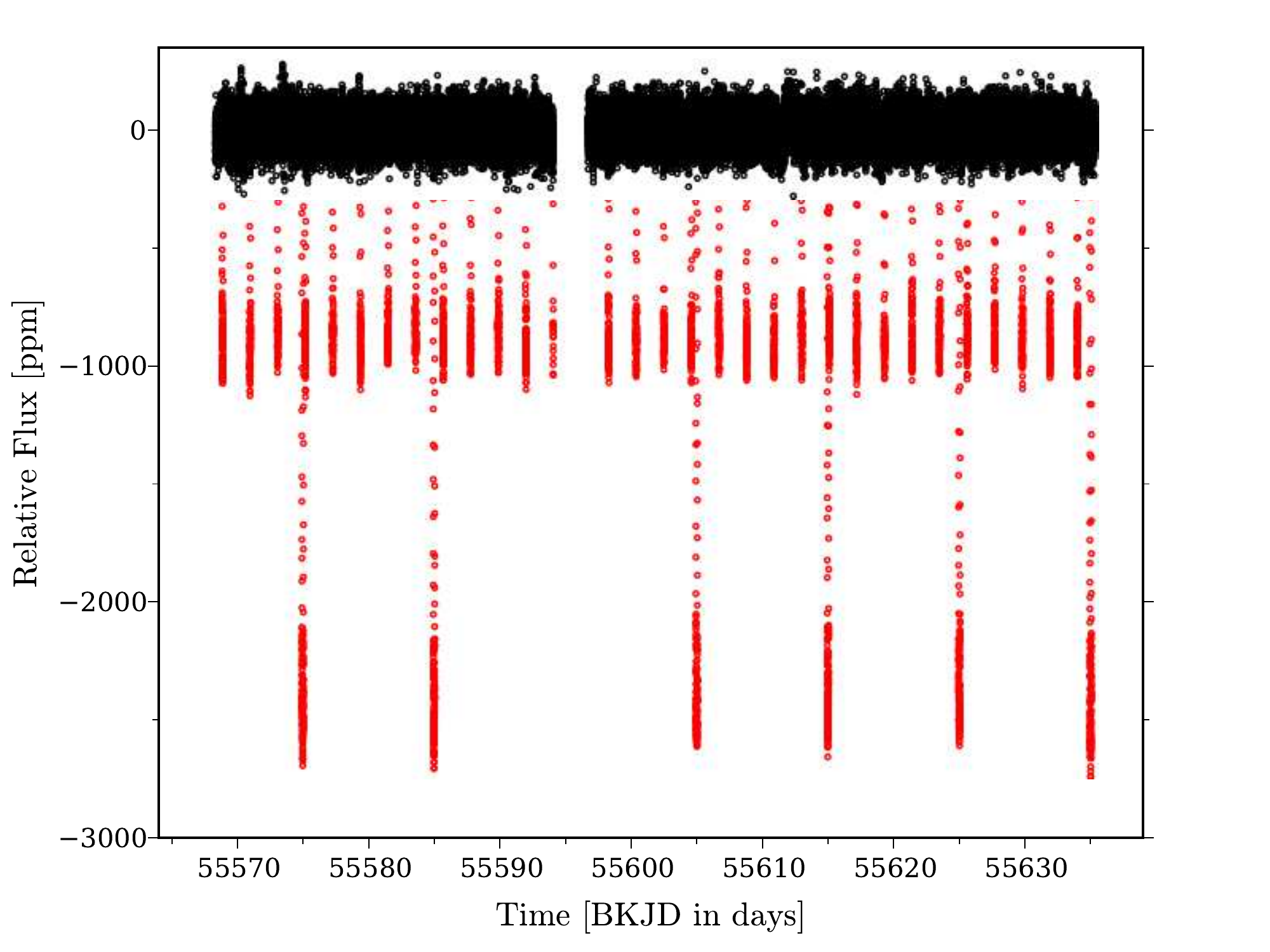}
	\caption{Segment of the time series of 16 Cyg A. In black we show the corrected original light curve, while in red the light curve is shown after the addition of a simulated transit signal.}%
	\label{fig:16CygA}%
\end{figure}

\begin{figure}%
	\centering
	\includegraphics[width=\columnwidth]{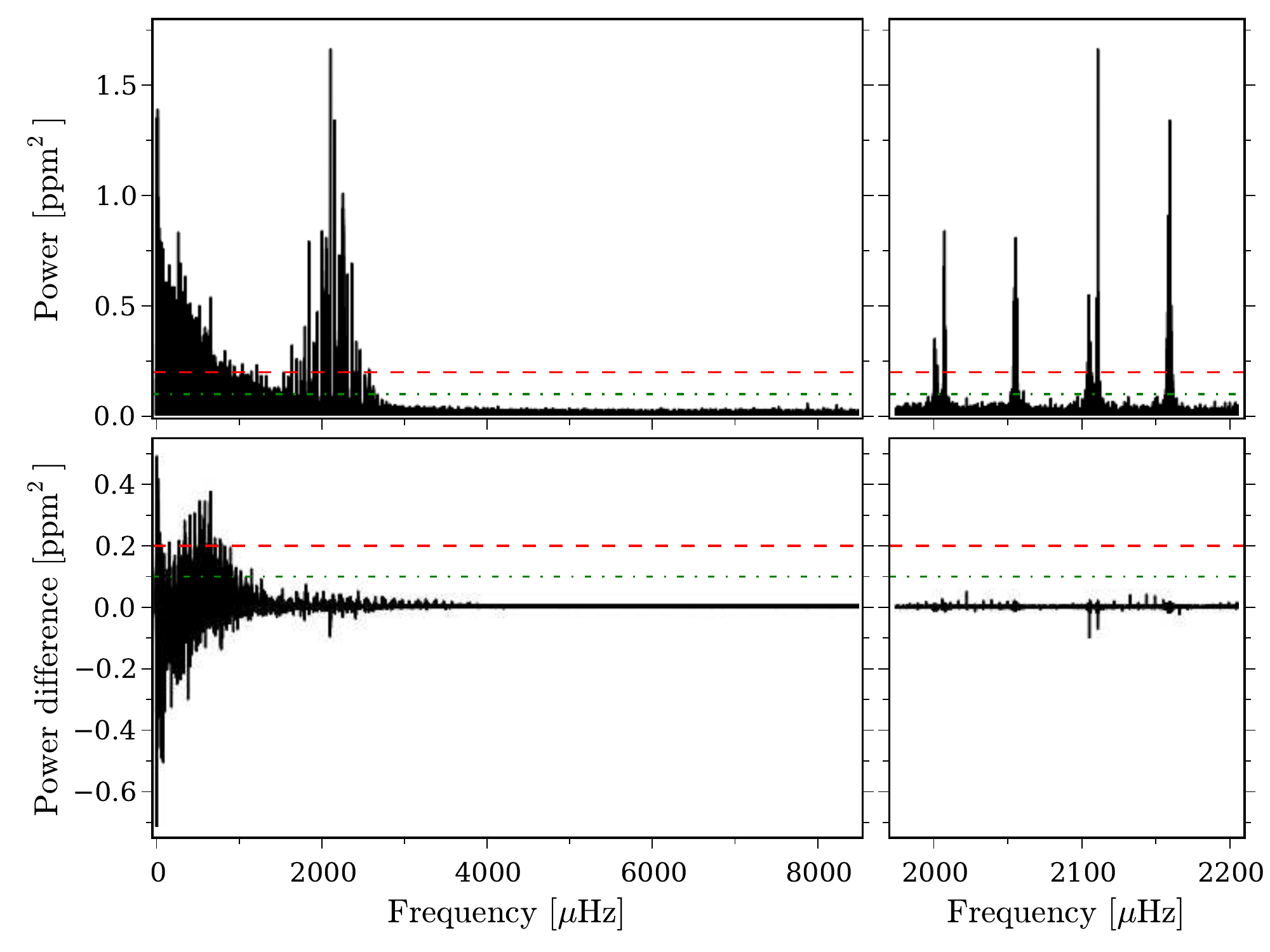}
	\caption{{\bfseries Top:} Corrected power spectrum of 16~Cyg~A, calculated from the lightcurve with an added transit signal. The left panels show the full power spectrum, while the right panels show a zoom on two of the central orders. {\bfseries Bottom:} Difference between the top power spectrum and the corrected original power spectrum. The two horizontal dashed lines are provided to guide the eye to the scale of the differences.}%
	\label{fig:16CygAps}%
\end{figure}

Finally, we want to test if the use of the phase-folded light curve might introduce high frequency noise to the power spectrum.
For this we use the time series of 16 Cyg A \citep[see, \eg,][]{2012ApJ...748L..10M,2014ApJ...782....2L}, which is one of the brightest targets in the \Kepler\ field-of-view and has a very high SNR (signal-to-noise ratio). First, the light curve was corrected using setting slightly modified from the default ones, in that a width of 1 day was used for the long filter rather than 3 days. Secondly, we multiplied this time series with a simulated transit signal\footnote{The transit signal was created using the Python modules \code{Bart} and \code{kplr}, which are being developed and supported by Dan Foreman-Mackey. See \url{http://dan.iel.fm/bart/} and \url{http://dan.iel.fm/kplr/}.}, see \fref{fig:16CygA}. The transit signal consisted of two planets, both of which have an impact parameter of $b=0.3$, and with periods of 10 and 2.1 days, and relative radii of $(R_p/R_\star)$ of $0.0316$ and $0.0223$ respectively. For the limb-darkening (LD) we adopted a quadratic law with LD-parameters of $\mu_1=0.4729$ and $\mu_2 = 0.1871$, corresponding to values from \citet[][]{2011A&A...529A..75C} using spectroscopic parameters for 16 Cyg A from \citet[][]{2009A&A...508L..17R}. Third, the total filter from the first step was multiplied back, such that the end result is the jump-corrected 16 Cyg A time series, but now with a transit signal. 
The filter was then run anew with the planetary periods specified.
In \fref{fig:16CygAps} we show the power spectra calculated for both the original and the time series with transits included.
The fact that 16 Cyg A has a very high SNR means that any added noise from the use of the filter on the added transit signal will stand out clearly.
We note that there is no systematic increase in the white noise level from the use of the phase curve in the correction or large residuals from the planet signal.

\section{KASOC data products}\label{sec:product}
All data products from the KASOC pipeline will become accessible in the KASOC database, at \url{http://kasoc.phys.au.dk/}. This includes the following products, most of which are described in detail above: 
\begin{enumerate}[(i)]
	\item Concatenated and corrected time series
	\item Power density spectra (weighted and un-weighted)
	\item Miscellaneous info in FITS headers
\end{enumerate}
In the following a description is given of the various outputs.
Here we also describe the data formats \etc of the data products. All output data will be provided in both FITS and ASCII formats. The FITS versions will contain all available information, while the ASCII files will only contain the most important information, but hence be easier to work with.

The files produced will be named in much the same way as currently done in the KASOC data base. The FITS files produced will carry names in the following format: 
``\code{kplr[KIC]\_[type]\_[slc or llc]\_v[version].fits}''. The ASCII files will follow the same naming convention, except of the extensions ``\code{dat}'' and ``\code{pow}'' for the time series and power spectrum respectively.

\subsection{Corrected time series}
The corrected time series as output from the KASOC filter will be accessible both in ASCII and FITS format. The ASCII files will, apart from a brief header, contain three columns with the timestamps in truncated barycentric Julian date ($\mathrm{BJD}-2400000$), corrected flux and associated error values respectively. The FITS files will contain two extension named ``\code{PRIMARY}'' and ``\code{TIMESERIES}''.
The first extension only holds a header which contains information on the star, the data used and the filter parameters used for the given star. See Table~\ref{tab:primaryheader} for a list of the information provided.
In particular, we would like to draw attention to the ``\code{FILEID}'' keyword. This number will uniquely identify the given dataset in the KASOC database and we encourage the users of the data to store this number together with any results arising from analysis of the data, as this will significantly improve traceability of the results.
The second extension contains a binary table with six columns. The first three of which are similar to the ASCII files: ``\code{TIME}'' which contain timestamps (in TBJD), ``\code{FLUX} which contains corrected relative fluxes (in ppm) and ``\code{FLUX{\_}ERR}'' which contain flux errors (in ppm).
In the fourth column, ``\code{FILTER}'', the final filter flux is given (in e$^-$/s). This will enable inspection of the exact effects the filter had on the data.
In the fifth column, ``\code{SAP{\_}QUALITY}'', we will keep the same bit values as found in the original data sets (see Appendix~\ref{sec:bit}). The last column ``\code{KASOC{\_}FLAG}'' will first of all describe the possible action taken for a given bit value, \ie, was the data point removed, was an offset correction made \etc. For a full list of possible flags see Table~\ref{tab:kasoc_bits}.

\begin{table}
\caption{Special keywords in the first/primary extension of the FITS files. }
\label{tab:primaryheader}
\centering
\begin{tabular}{lp{6cm}}
\toprule
Keyword & Description \\
\midrule
\code{KEPLERID} & Kepler Input Catalog (KIC) identifier. \\
\code{KOI} & Kepler Object of Interest  (KOI) identifier. \\
\code{FILEID} & Unique ID of the data file in the KASOC database. \\
\code{OBSMODE} & Kepler observing mode (LC/SC). \\
\code{VERSION} & Version of the data. \\
\code{VERPIXEL} & Version of pixel mask algorithm used. \\
\code{VERCORR}  & Version of KASOC filter used. \\
\code{QUARTERS} & Kepler observation quarters used in the file. \\
\code{KF{\_}MODE} & Filter operation mode. \\
\code{KF{\_}LONG} & Long timescale used ($\tau_\mathrm{long}$) in days. \\
\code{KF{\_}SHORT} & Short timescale used ($\tau_\mathrm{short}$) in days. \\
\code{KF{\_}SCLIP} & Sigma clipping. \\
\code{KF{\_}TCLIP} & Turnover clip ($\mu_\mathrm{to}$). \\
\code{KF{\_}TWDTH} & Turnover width ($\sigma_\mathrm{to}$). \\
\code{KF{\_}PSMTH} & Phase smooth factor. \\
\code{NUM{\_}PER} & Number of periods used. \\
\code{PER{\_}$n$}  & Values of periods used.\newline$n\nobreak=\nobreak1,2\dots,$\code{NUM\_PER}. \\
\bottomrule
\end{tabular}
\end{table}

\begin{table}
\centering
\caption{Flags currently available in the ``\code{KASOC{\_}FLAGS}'' column in the FITS files.}
\label{tab:kasoc_bits}
\begin{tabular}{rrp{5.7cm}}
\toprule
Bit & Value & Description \\
\midrule
  - &    0 & Good data point. \\
  1 &    1 & Point removed prior to filter as flagged bad data point. \\
  2 &    2 & Timestamp used to correct jump using constant offset. \\
  3 &    4 & Timestamp used to correct jump using linear function. \\
  4 &    8 & Data removed as a result of sigma clipping. \\
  5 &   16 & Possible transit structure ($c>0.5$ and $x_\mathrm{short} < x_\mathrm{long}+x_\mathrm{transit}$). \\
\bottomrule
\end{tabular}
\end{table}

\subsection{Power density spectra}
The power density spectra files will overall follow the same structure as the files for the corrected time series. Two separate files will be provided for the weighted and un-weighted power spectra, but using the same file structure. The ASCII files will, apart from a short header, simply contain two columns with frequencies in $\rm \mu Hz$ and power density in $\rm ppm^2 / \mu Hz$.
The FITS files will contain two extensions named ``\code{PRIMARY}'' and ``\code{POWERSPECTRUM}''. The first one will, as for the lightcurve files, contain header information uniquely identifying the star and the data file. The second extension contains a binary table with two columns. The first column (``\code{FREQUENCY}'') contain frequencies in $\rm \mu Hz$ and the second column (``\code{PSD}'') contain the associated power spectral densities in $\rm ppm^2 / \mu Hz$.


\section{Discussion and future development}\label{sec:dis}
In this paper we have presented a prescription on how to automatically pre-process solar-like asteroseismic data from the {\Kepler}/K2 space craft in an efficient way that can easily be applied to a large number of stars. We have demonstrated that the filter does a very good job of removing unwanted instrumental and planetary signals from the time series, resulting in better datasets for asteroseismic analyses.

We will not argue that the procedures presented here will be optimal in all circumstances, but for most targets it will significantly improve the data for asteroseismic usage.
In individual cases, a detailed filtering of the data could possibly provide even better results, but in many cases this would constitute simple modifications of the presented filter. Therefore, the software developed for the KASOC Filter allows for simple changes of the filter parameters on a per-star basis and the possibility of expanding and modifying the procedures, given additional knowledge of the particular target.
For example, if we would be in possession of a full transit model from a detailed fit to the planetary transits this can be fed directly into the filter by replacing the calculation of $x_\mathrm{transit}$ with the actual transit model.
However, this is not routinely done as the fits of the full transit model are non-trivial and relatively computational expensive to obtain. This often results in fits not utilizing all available data, and they are rarely updated with newer data.
Another issue is that standard planetary transit fits \citep[see, \eg,][]{MandelAgol2002} do not take all physical effects caused by the planet into account, like \eg planet phases, ellipsoidal variations, and relativistic beaming. Our filter on the other hand does not assume any physical model. The only assumption is that all features present in the time series with the same period as the planet are due to the planet and should therefore be removed.
We will not argue that the presented method is more ``correct'' than doing a real detailed physical model of the planetary transits or other features present in the data, assigning real physical phenomenon to each of them and correcting them using real descriptions of the physics behind each phenomenon.
However, this would cause a conflict with our design criterion that the filter should be automated and computationally fast. We have nonetheless integrated the possibility to apply a real transit model to the time series if such a model should exist. This could be applied in special cases where the described procedures would do a less than satisfactory job.
One situation where a full transit model could be beneficial could be the case of multi-planet systems showing large transit timing variations (TTVs) due to the gravitational interactions between the planets.

A few points that we are considering as possible expansions or advancements of the procedures presented here includes the following:
\begin{itemize}[$\circ$]
	\item Correct for unstable telescope pointing by decorrelating the photometric noise with the measured position of the star on the CCD \citep[see \eg][]{Vanderburg2014}. This will be particularly important in the context of the K2 mission.
	\item Use weighted medians taking into account measurement errors coming from the pixel extraction. For now, these are assumed to be constant.
	\item Allow for extraction of flux via fit of \Kepler point spread functions \citep[][]{2010ApJ...713L..97B}, \eg\ by use of the PyKE software package \citep[][]{2012ascl.soft08004S}.
	\item Test the use of co-trending basis vectors (CBV) for removal of instrumental effects \citep[see][]{KeplerDataHandbook}.
	\item Test gap filling techniques, such as \eg inpainting \citep[][]{2010arXiv1003.5178S}, to reduce noise in the power spectrum \citep[see also][]{garcia_to_come}.
	\item An automated selection of the optimum filter widths for individual stars. This could possibly be obtained by use of \emph{cross validation}.
\end{itemize}

Even though the procedures presented here are specifically designed for the \Kepler mission, most of it could easily also be applied to datasets from other sources, like the upcoming TESS mission \citep{TESS}.

\section*{\textbf{Acknowledgements}}
The authors wish to thank Hans Kjeldsen for developing the initial ideas for the filter and Steven Bloemen for providing the code for redoing pixel masks.
The authors would like to thank Jens Jessen-Hansen and Vincent Van Eylen for many useful discussions, and for tests of the filter.
We also thank Yvonne Elsworth for reading and commenting on earlier drafts of this paper.
Funding for the Stellar Astrophysics Centre (SAC) is provided by The Danish National Research Foundation. The research is supported by the ASTERISK project (ASTERoseismic Investigations with SONG and Kepler) funded by the European Research Council (Grant agreement no.: 267864).
RH acknowledges support from the UK Science and Technology Facilities Council.
The research leading to these results has received funding from the European Community's Seventh Framework Programme (FP7/2007-2013) under grant agreement no. 269194.
MNL would like to thank Dennis Stello and his colleagues at the Sydney Institute for Astronomy (SIfA) for their hospitality during a stay where some of the presented work was done.

This research has made use of the following web resources: NASAs Astrophysics Data System Bibliographic Services (adswww.harvard.edu); arxiv.org, maintained and operated by the Cornell University Library.

\appendix

\section{Bit values}\label{sec:bit}
In Table~\ref{tab:bits} we give the bit values for the different effects included in the ``Quality'' entry of the \Kepler FITS files.
We refer the reader to \citet[][]{kepman} for more information on the different phenomena included.
Note, that if a specific datum has been subject to more than one artefact, say \eg that a cosmic ray was detected in the optimal aperture pixel in addition to there being an attitude tweak, then the value of the ``Quality'' entry for the data point would be 129 (128+1), and so some care must be exercised in decomposing the value into its constituent bit components.

\begin{table*}
\centering
\caption{Flags given in the original \Kepler data files and the action taken for each of them in the filter. ``Remove'' refers to the datum being removed (set to NaN) by the filter. ``Correct'' means that the timestamp is used to correct a jump in the time series. ``Ignore'' means that the filter simply ignores the flag. See \citet[][]{kepman} for details.}
\label{tab:bits}
\begin{tabular}{rrll}
\toprule
Bit & Value & Description & Action \\
\midrule
  1 &    1 & Attitude Tweak. & Correct, Remove. \\
  2 &    2 & Spacecraft in Safe Mode. & Ignore. \\
  3 &    4 & Spacecraft in Coarse Point. & Ignore. \\
  4 &    8 & Spacecraft in Earth Point. & Ignore. \\
  5 &   16 & Reaction wheel zero crossing. & Ignore. \\
  6 &   32 & Reaction wheel desaturation event. & Remove. \\
  7 &   64 & Argabrightening detected across multiple channels. & Ignore. \\
  8 &  128 & Cosmic Ray in Optimal Aperture pixel. & Ignore. \\
  9 &  265 & Manual Exclude. The cadence was excluded because of an anomaly. & Remove. \\
 10 &  512 & Not in use. & Ignore. \\
 11 & 1024 & Discontinuity corrected between this cadence and the following one. & Correct. \\
 12 & 2048 & Impulsive outlier removed after cotrending. & Ignore. \\
 13 & 4096 & Argabrightening event on specified CCD mod/out detected. & Remove. \\
 14 & 8192 & Cosmic Ray detected on collateral pixel row or column in optimal aperture. & Ignore. \\
\bottomrule
\end{tabular}
\end{table*}

\section{Median filter implementation}\label{sec:median}
The standard way to run a median filter is to, for a given data point, calculate the median value, using some build-in function, in a window comprising a given number of neighbouring data points - then one moves to the next point and calculates the median value anew and so forth. However, this is very inefficient as the build-in median function for each step need to perform a sorting of the points contained within the set window. In our implementation we first of all make sure that the window comprises a odd number of point, whereby the median is given by the value at the midpoint index of the sorted values - and the value of this index does not change if the window for each step in the moving of the filter comprise the same (odd) number of points. In the first step of the moving median, \ie in the calculation for the first point in the time series, we sort the values comprised by the set window and find the mid index. When moving the filter to the next data point we do, however, not throw away our sorted values from the first step. Instead, we merely remove from the sorted values the data point that is now no longer within the window (the values are still sorted), and then add the new data point that has come within the window and place it at the appropriate position among the sorted values. With a bit of housekeeping the removal of one point and addition of another to the sorted values can be done very easily, and the heavy reduction of operations greatly enhance the speed of the moving median. We do by no means argue that this is the most efficient way to implement a moving median filter, but it does result in a great improvement in speed and the coding is quite simple.   

\bibliography{bibliography}
\end{document}